\documentclass[11pt,twoside,dvips]{article}
\usepackage{verbatim}
\usepackage{amssymb}
\usepackage{amsfonts}
\usepackage{amsmath}
\usepackage{pifont}
\usepackage[spanish,english]{babel}
\usepackage[pdftex]{graphicx}
\usepackage{verbatim}
\usepackage{epsfig}
\usepackage{bm}
\usepackage{longtable}
\usepackage{fancyhdr}
\begin{document}

\fancypagestyle{plain}{%
\fancyhf{}%
\fancyhead[LO, RE]{XXXVIII International Symposium on Physics in Collision, \\ Bogot\'a, Colombia, 11-15 september 2018}}

\fancyhead{}%
\fancyhead[LO, RE]{XXXVIII International Symposium on Physics in Collision, \\ Bogot\'a, Colombia, 11-15 september 2018}

\title{Millepede alignment of the Belle 2 sub-detectors after first collisions}
\author{Tadeas Bilka$\thanks{%
e-mail: bilka@ipnp.mff.cuni.cz}$, Jakub Kandra for the Belle II Collaboration, \\ Faculty of Mathematics and Physics, Charles University, \\ 121 16 Prague, Czech Republic}
\date{}
\maketitle


\begin{abstract}
The Belle II detector at the SuperKEKB accelerator observed the first collisions in April this year. Until mid-summer, the first commissioning run uses a reduced version of the full vertex detector. Nevertheless, this phase is an excellent opportunity to improve and test the alignment and calibration procedures being prepared for the first physics runs starting in spring 2019.

The procedure presented is based on Millepede II tool to solve the large minimization problem emerging in the track-based alignment and calibration of the pixel and strip detectors, the drift chamber or the muon system. The first alignment of the vertex detector was performed quickly after the first collisions and further improvements are expected with more data and with inclusion of other sub-detectors into the procedure.

This contribution will show overview and status of the Millepede alignment and calibration procedure of the Belle 2 sub-detectors,  after first collisions and the plans for full physics run.
\end{abstract}

\section{Introduction}
The Belle II experiment is an upgrade of the Belle detector and KEKB accelerator at KEK laboratory (Tsukuba, Japan) \cite{TDR, belle2}. The new SuperKEKB accelerator with 40 times higher design luminosity than KEKB is an asymmetric $e^+/e^-$ collider at $E_{CM} = m_{\Upsilon(4S)}$ (10.58 $GeV/c^2$). The experiment will study CP violation and heavy flavour physics in B, D or tau decays and probe the physics beyond the standard model.

The Belle II detector aims to record 50 $\mathrm{ab^{-1}}$ data sample. To cope with larger backgrounds and to reach the desired physics performance, from vertexing to particle identification, the detector undergone substantial upgrades.

High target performance of the detector requires fast and reliable calibration and alignment procedures.
We present a single method for track-based alignment of the vertex, tracking and muon detector with a global approach of Millepede II, advanced track model and material treatment. As the experiment entered its commissioning phase, we also show a small example of some of the initial results obtained with the described procedures applied on the first collision data and the vertex detector.

\begin{figure}
\centering
\includegraphics[width=10cm,clip]{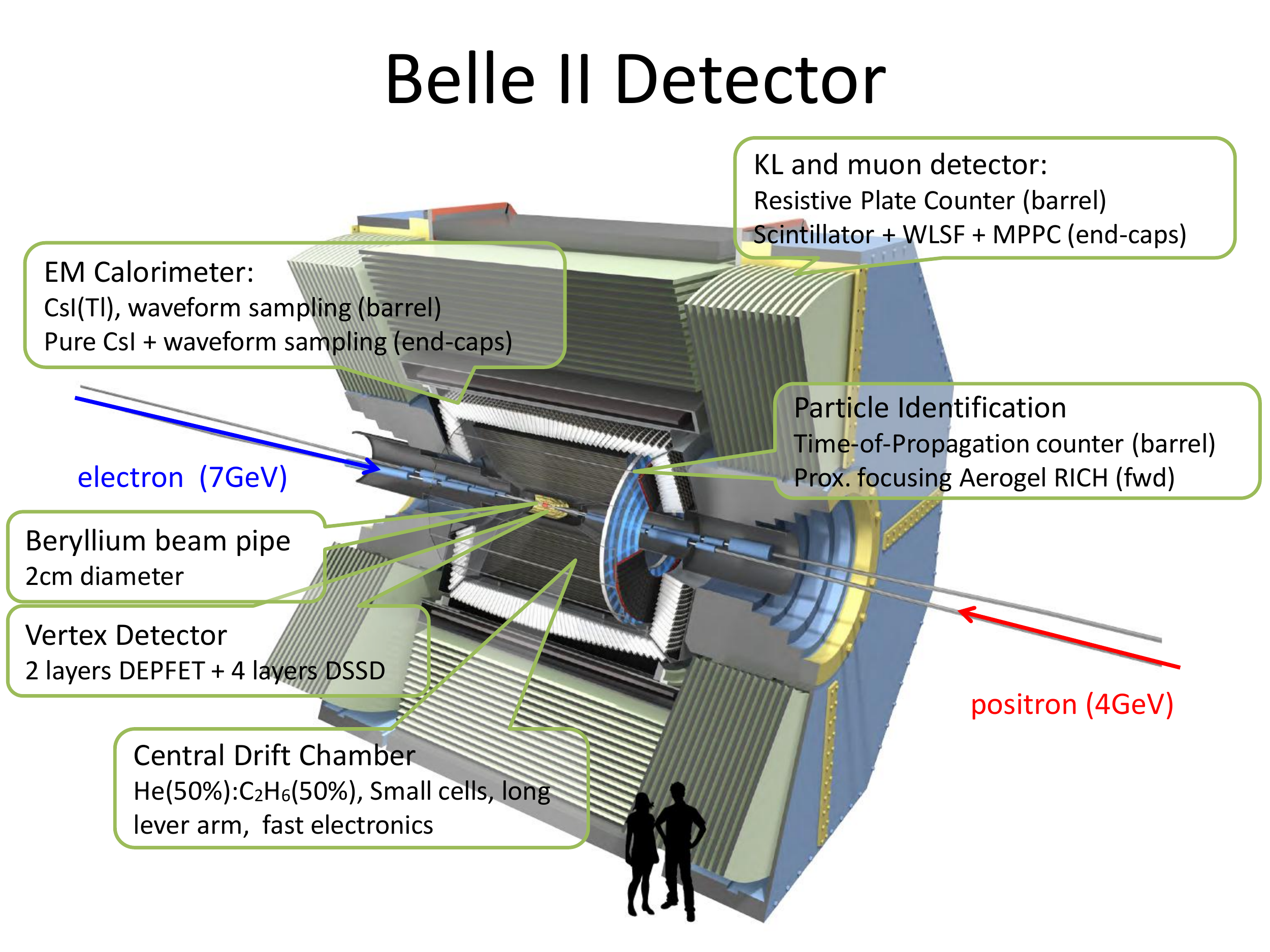}
\caption{Cross-section of the Belle II detector with its sub-detector systems: vertex detector, drift chamber, particle identification systems, electromagnetic calorimeter and the muon system which also serves as a return yoke for the 1.5 T superconducting magnet. Main new and/or upgraded features of each sub-detector are highlighted in the captions.}
\label{fig-BelleII}       
\end{figure}


\section{Belle II sub-detectors}
The cross-section of the Belle II detector, showing all its sub-detectors is shown in Fig. \ref{fig-BelleII}. We concentrate here on the sub-detectors involved in the procedure.

For precise vertex reconstruction, the innermost sub-detector, the pixel detector employs about 8~million pixels based on DEPFET (Depleted P-channel Field Effect Transistor) technology. The sensors are organized in two layers (first at radius 14 mm from the beam line), with 8 sensors (50 $\mathrm{\mu m}$ x 55 -- 60 $\mathrm{\mu m}$ pixels) in the 1st layer and 12 in the second (50 $\mathrm{\mu m}$ x 70 -- 85 $\mathrm{\mu m}$ pixels). All sensors' active areas are thinned to 75 $\mathrm{\mu m}$.

The silicon vertex detector completes the vertexing device. It allows to use extrapolation to PXD for a significant data reduction or to reconstruct decays of $K_S$'s outside of PXD. It is composed from 187 double-sided silicon strip sensors (300 -- 320 $\mathrm{\mu m}$ thickness) with fast readout in four layers.
1st layer is at radius of 38 mm (strip pitch z/$R-\phi$: 50 / 160 $\mathrm{\mu m}$), layers 2/3/4 are placed at radii 80/115/140 mm (strip pitch z/$R-\phi$: 75 / 240 $\mathrm{\mu m}$) and are equipped with slanted trapezoidal sensors in the forward part.

For charged particle momentum reconstruction and dE/dx particle identification, Belle II uses
a central tracking device, the central drift chamber (CDC). CDC is a 2 m long He-$C_2$$H_6$ gas wire chamber, with inner / outer radius of 16 cm / 113 cm and 14336 sense wires (W + Au plating) arranged in 56 alternating axial and stereo layers, with small cell chamber near the interaction point.

The outermost sub-detector, barrel (BKLM) and end-cap (EKLM) $K_L$ and muon detector,
consists of alternating detection elements and iron plates (used also for magnetic flux return). There are
14 layers of segments with scintillator strips with Si-PMs readout in EKLM (and two innermost layers of BKLM) and
13 layers of resistive plate counter modules in BKLM (two in each iron gap).

\section{Alignment Tools}

\subsection{Software Implementation}
The methods for detector alignment are implemented within the Belle 2 Analysis Software Framework (basf2), a modern modular HEP framework (C++11/Python), developed for the experiment \cite{basf2}. The framework covers almost all software related tasks, from data acquisition and simulation to online/offline reconstruction and physics analysis.

The Calibration and Alignment  Framework (CAF) is a basf2 package designed to simplify and automatize calibration work-flows. It deals with the management of many different calibration algorithms, dependencies among them, data aggregation, job submission (local, batch...), iterations or communication with a database. The package is further extended by an automated calibration package b2cal (using Apache Airflow).

The Millepede II algorithm, discussed below, is fully integrated in CAF and profits from tight integration of several tools within basf2. The framework integrates the generic track fitting toolkit GENFIT2 \cite{genfit}, providing all necessary ingredients for track fitting, from Runge-Kutta track extrapolation in in-homogeneous magnetic field to treatment of measurements of arbitrary dimension (pixel, strip, wire, dE/dx ...) via the generic formalism of virtual measurement planes. We integrated the General Broken Lines (GBL) into the GENFIT2 toolkit.

GBL is a track model and linear least squares re-fit with proper
description of multiple scattering \cite{gbl}. A particle trajectory is constructed from points with a measurement
and/or a thin scatterer (source of track slope variance). Fit parameters also include kink angles at scatterers, common curvature correction or drift time corrections (CDC). The global linearized $\chi^2$ fit by minimization of residuals and kinks yields full covariance matrix of the track (including correlations from multiple scattering).
The GBL trajectory is constructed from reconstructed track or multiple tracks from a common decay for composite trajectories. The reconstructed track is re-extrapolated to integrate material distribution and calculate moments of radiation length along particle trajectory. The material between each two measurement planes is then represented by doublet of thin scatterers.

\subsection{Millepede II Alignment}
Millepede II \cite{millepede_article} is a tool designed to solve certain linear least squares problems with very large number of parameters. The problem is solved by minimization performed with respect to all global (calibration) and local (track) parameters, but only global parameters are determined. All correlations are kept in the solution, which is exact for linear problems and can be iterated to cope with non-linearities.

%
%

Millepede II is integrated into CAF and basf2. The input consists of trajectories constructed by GBL from single tracks or e.g. mass-and-vertex constrained two-body decays. Rich topology of the samples (from beam data to cosmics with or without magnetic field) and additional physics constraints are crucial to reduce possible systematic distortions (weak modes) to which the $\chi^2$ minimization is insensitive. The considered alignment parameters are six rigid body paremeters (3 shifts + 3 rotations in local frame of the rigid body) for each sensor (PXD, SVD), module (BKLM) or segment (EKLM). In addition, the alignment of higher-level structures (e.g. ladders and half-shells of PXD and SVD) is considered. For CDC, two shifts and one rotation is considered on both end-plates for all layers, with possible extension up to wire-by-wire alignment. Finally, as the position of the primary vertex can enter some constraints, its alignment is implemented as well.
 
\section{First Collisions in Commisioning Phase 2}
The Belle II experiment start-up is divided to three phases:
\begin{itemize}
\item
Phase 1, devoted to SuperKEKB accelerator commissioning without collisions.
\item
Phase 2, aimed at SuperKEKB and Belle II detector (background) commissioning with collision data. This phase uses a reduced vertex detector (one section, 18 sensors) and dedicated background radiation detectors. The first collisions were recorded on 26th April 2018 and the this first data taking period finished on 17th July 2018.
\item
Phase 3, starting in spring 2019 -- physics run with full vertex detector.
\end{itemize}

\begin{figure}[h]
\centering
\includegraphics[width=12cm,clip]{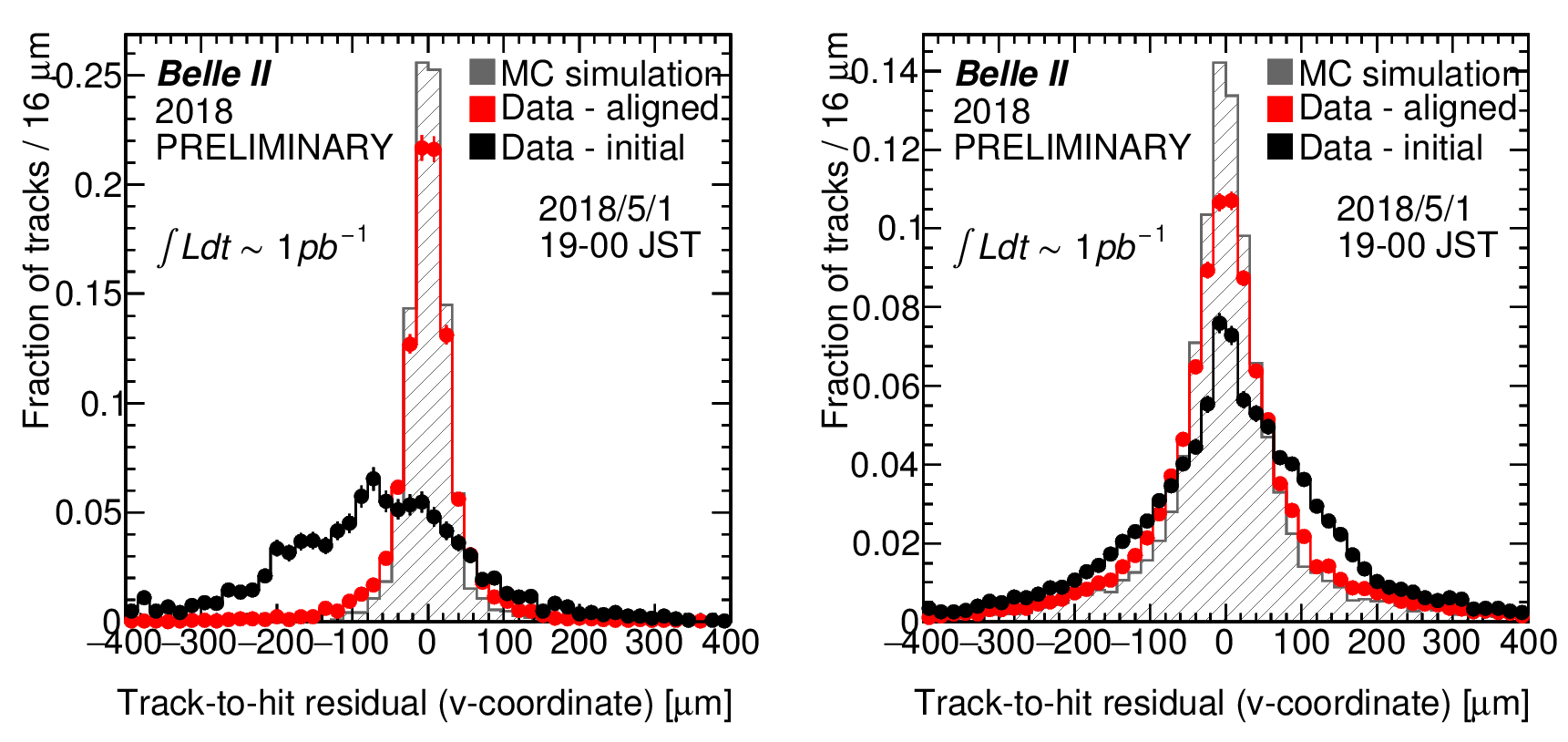}
\caption{
Histograms  from  alignment  DQM  modules  show  track-to-hit  residuals  from  all  PXD
(left) and SVD (right) sensors after standard reconstruction and track fitting is performed.
Distributions  are  shown  for  the  residuals  of  the  local  v-coordinate, parallel to z/beam axis.  The data from some of the first collisions
are used to illustrate the improvement of the position resolution before (black) and after
(red)  the  first  preliminary  alignment  is  computed  and  used  in  reconstruction. All 108 parameters (18 sensors $\times$ 6 rigid body parameters, CDC fixed as reference) are determined simultaneously.  A  sample
of 20k charged particles originating near IP (electrons, positrons, protons) is generated to check the correspondence of the data and MC simulation (grey).  All compared histograms
are normalized.
}
\label{fig-VXDAlignment}       
\end{figure}

\begin{figure}
\centering
\includegraphics[width=16cm,clip]{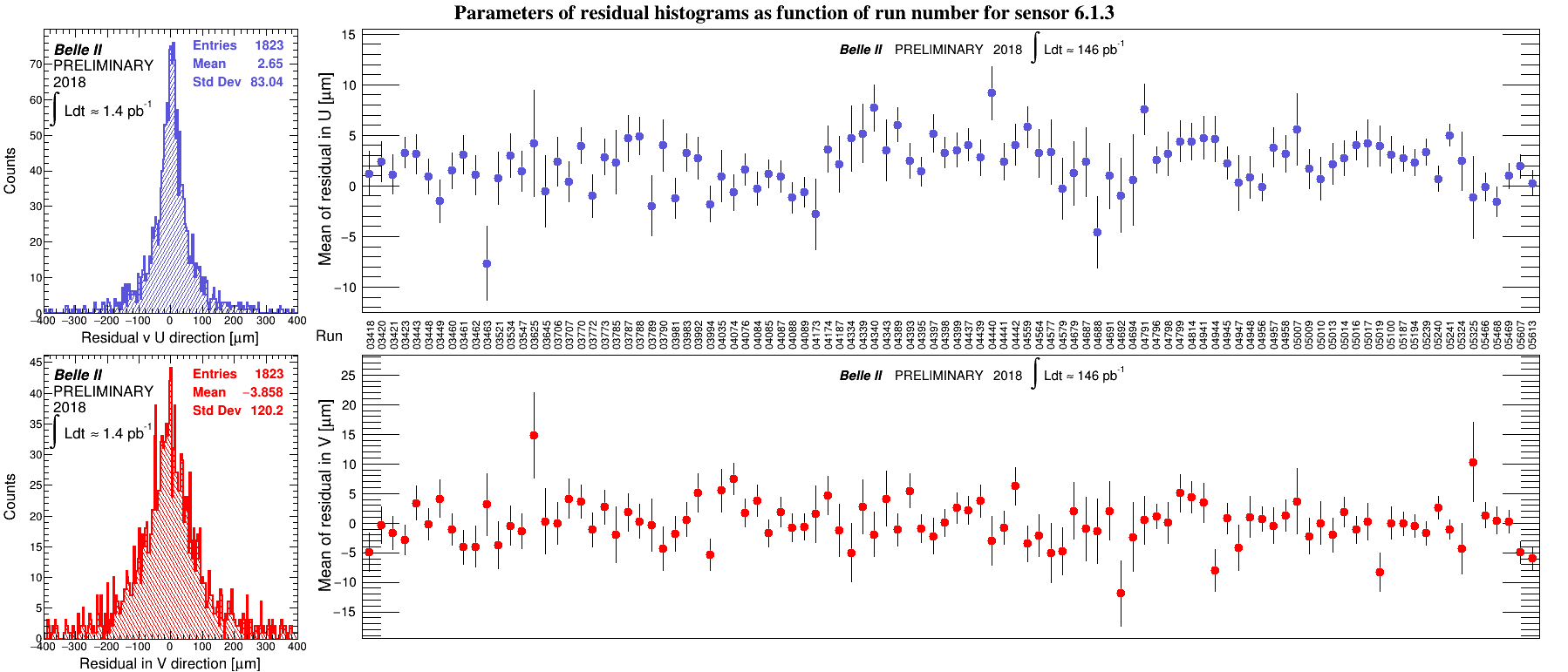}
\caption{The histograms of the track-to-hit residuals of tracks used for alignment (left) and evolution of the mean track-to-hit residual over the last 100 runs of phase 2 (right) for a single sensor (3rd SVD sensor in last layer) in local $u$-coordinate (global $R-\phi$ direction) (top) and local $u$-coordinate (global $z$ direction) of the sensor (bottom).}
\label{fig-stability}       
\end{figure}

The first results were obtained with the vertex detector alignment hours after the first collisions (see Fig. \ref{fig-VXDAlignment}) and validated over the period of Phase 2 (see Fig. \ref{fig-stability}). The alignment of other sub-detectors and the possibility to perform the alignment of all the sub-detectors simultaneously is currently studied with the phase 2 data.

\section{Conclusions}
The pixel and strip detectors, the drift chamber and the muon detector and primary beam spot alignment are integrated into GBL fitting and Millepede II alignment and calibration.
The calibration infrastructure was successfully tested with the first collision data in Phase 2.
Vertex detector alignment, as the most mature and well-tested part, is showing good performance on data. The extension of the method to the drift chamber and muon system is under extensive testing with Phase 2 data.
The tools are getting ready for Phase 3 with additional features, which are under various level of development or validation. These include usage of vertex and mass constrained decays, alignment of sensor deformations or additional calibration parameters for CDC.
\section*{Acknowledgment}

The study was supported by the Charles University (project GA UK No. 404316), Ministry of education, youth and sports (Czech Republic) and MSCA-RISE project JENNIFER (EU grant n.~644294).


\end{document}